\documentclass[%
 aip,
 bibnotes,
 nofootinbib,
 amsmath,amssymb,
 reprint,%
]{revtex4-1}
\usepackage{graphicx}
\graphicspath{ {figures/} }

\usepackage{dcolumn}
\usepackage{bm}
\usepackage{hyperref}
\usepackage{notes2bib}
\bibnotesetup{
note-name = ,
use-sort-key = false
}

\begin{document}

\title{Single Photon Sources with Near Unity Collection Efficiencies by Deterministic Placement of Quantum Dots in Nanoantennas}

\author{Hamza Abudayyeh}
\author{Boaz Lubotzky}
\affiliation{Racah Institute of Physics, The Hebrew University of Jerusalem, Jerusalem 9190401, Israel}
\affiliation{The Center for Nanoscience and Nanotechnology, The Hebrew University of Jerusalem, Jerusalem 9190401, Israel}

\author{Anastasia Blake}
\author{Jun Wang}
\author{Somak Majumder}
\author{Zhongjian Hu}
\author{Younghee Kim}
\author{Han Htoon}
\affiliation{Materials Physics \& Applications Division: Center for Integrated Nanotechnologies, Los Alamos National Laboratory, Los Alamos, New Mexico 87545, United States}

\author{Riya Bose}
\author{Anton V. Malko}
\affiliation{Department of Physics, University of Texas at Dallas, Richardson, Texas 75080, United States}

\author{Jennifer A. Hollingsworth}
\email{jenn@lanl.gov}
\affiliation{Materials Physics \& Applications Division: Center for Integrated Nanotechnologies, Los Alamos National Laboratory, Los Alamos, New Mexico 87545, United States}

\author{Ronen Rapaport}
\email{ronenr@phys.huji.ac.il}
\affiliation{Racah Institute of Physics, The Hebrew University of Jerusalem, Jerusalem 9190401, Israel}
\affiliation{The Center for Nanoscience and Nanotechnology, The Hebrew University of Jerusalem, Jerusalem 9190401, Israel}
\affiliation{The Applied Physics Department, The Hebrew University of Jerusalem, Jerusalem 9190401, Israel}
\date{\today}

\footnote{Authors to whom correspondence should be addressed: ronenr@phys.huji.ac.il, jenn@lanl.gov}
\begin{abstract}
Deterministic coupling between photonic nodes in a quantum network is an essential step towards implementing various quantum technologies. 
The omnidirectionality of free-standing emitters, however, makes this coupling highly inefficient, in particular if the distant nodes are coupled via low numerical aperture  (NA) channels such as optical fibers. This limitation requires placing quantum emitters in nanoantennas that can direct the photons into the channels with very high efficiency. 
Moreover, to be able to scale such technologies to a large number of channels, the placing of the emitters should be deterministic. 
In this work we present a method for directly locating single free-standing quantum emitters with high spatial accuracy at the center of highly directional bullseye metal-dielectric nanoantennas.
We further employ non-blinking, high quantum yield colloidal quantum dots (QDs) for on-demand single-photon emission that is uncompromised by instabilities or non-radiative exciton recombination processes.
Taken together this approach results in a record-high collection efficiency of 85$\%$ of the single photons into a low NA of 0.5, setting the stage for efficient coupling between on-chip, room temperature nanoantenna-emitter devices and a fiber or a remote free-space node without the need for additional optics.

\end{abstract}

\maketitle

\section{Introduction}
The physical realization of the photonic building blocks of future quantum technology and communication applications relies on the development of highly efficient and reliable quantum emitters at photonic nodes. \cite{Lounis2005,Aharonovich2016Solid-stateEmitters,Rodt2020DeterministicallySources,Laucht2012ASource}.
Solid-state emitters operating at cryogenic temperatures are currently the state-of-the-art in terms of single photon performance.  \cite{Gazzano2013BrightPhotons,Ding2016,Somaschi2016NearState,Liu2018HighPhotons,Wang2019TowardsMicrocavities}
Although the technology is relatively well established for these devices, there remain significant obstacles with respect to scaling and network integration. 
Furthermore operation of the single-photon source would ideally not be limited to ultra-low temperatures.
Room-temperature sources, on the other hand, face different challenges.
Like their cryogenic counterparts, they are characterized by isotropic emission, which necessitates that the emitter is coupled with a light-directing nanoantenna. 
In addition, however, they are often characterized  by: (a) relatively broadband emission which impacts coupling strategies, leading to stringent fabrication requirements and can limit the practical useful brightness that can be collected into photonic channels ; (b) low single-photon purity; and (c) unstable emission. 
Thus, new approaches are needed to achieve the combined attributes of pure, stable, room-temperature operation; scalable, facile and targeted fabrication, and high efficiency in the anticipated device architectures. 

The isotropic nature of the emitters can been addressed by modifying the photonic environment near the quantum emitter.
This is achieved by coupling the emitter to metallic or dielectric nanostructures that serve as plasmonic or optical antennas or resonators.
Metallic and plasmonic nanostructures such as metal  nanoparticles  (MNPs) \cite{Yuan2009AntibunchingDots,Menagen2009AuMechanisms,Ma2010FluorescenceNanoparticles,Ji2015Non-blinkingResonator} ,  plasmonic  patch  antennas \cite{Belacel2013ControllingAntennas,Hoang2016UltrafastNanocavities,Dhawan2020ExtremeAntennas} , and circular bullseye plasmonic nanoantennas \cite{Yi2014BeamingAntenna,Harats2014,Andersen2018HybridCollection} generally have low mode volumes and quality factors enabling emission modification over a broadband spectrum. 
However, these structures have stringent distance requirements for locating the emitter relative to the antenna, and the metal can have the unwanted side effect of increasing the rate of non-radiative recombination processes. 
On the other hand, dielectric structures, such as microcavities \cite{Skolnick1998StrongStructures,Reithmaier2004StrongSystem,Press2007PhotonRegime,Reitzenstein2007AlAs-GaAs150000,Ding2016,Somaschi2016NearState} and photonic crystals \cite{Kaniber2008InvestigationNanocavities,Liu2018HighPhotons} feature high radiative enhancement factors and low-loss, but generally  come  with  a  narrow  frequency  bandwidth,  which  is  usually unsuitable for broadband room-temperature quantum emitters.

\begin{figure}[t]
  \centering
    \includegraphics{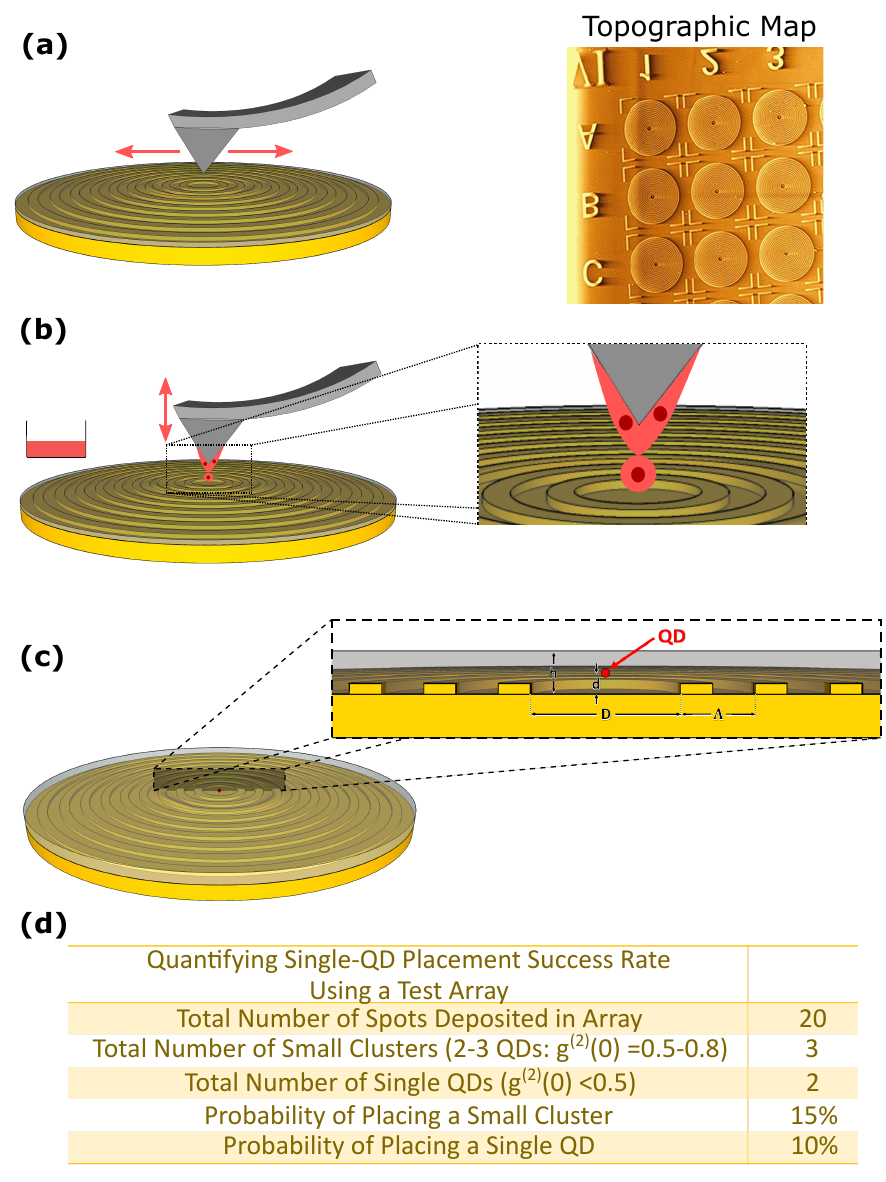}
    \caption{ Sketch of hybrid metal-dielectric bullseye structure used in this study and the DPN placement technique. (a) An AFM scan of the area is conducted (the right panel shows an actual image of the scanning of an antenna array), then (b) the tip is wetted and a droplet is placed at the center. After the fabrication process is finalized the resulting structure is shown in (c). (d) Statistics on placement probabilities. \cite{supp} \label{fig: intro} }
\end{figure}

To capture the distinct benefits of both metal and dielectric structures, we previously demonstrated the use of hybrid metal-dielectric bullseye nanoantennas \cite{Livneh2015EfficientNanoantenna} for efficient collimation of radiation from nano-emitters.
In this hybrid structure (see Fig \ref{fig: intro}c) the photon source located at the center of the bullseye emits into a dielectric layer that acts as a slab waveguide guiding the light radially outward towards the circular gratings. 
The parameters of the antenna can be tuned in such a way that in the far-field the interference between the various diffracted waves occurs only at low angles thus resulting in a highly directional photon stream. 
In such a design, the emitter can be located inside a low loss dielectric medium, at a large distance from the metal, thus avoiding metal induced losses yet still producing very high directionality in a broad spectral range.
Following our first demonstration with many emitters, we showed that a single colloidal quantum dot (QD) could be coupled to such an antenna and result in a device with highly directional single-photon emission.  \cite{Livneh2016,Harats2017DesignEmission}.

However, a major outstanding challenge remains that severely limits the performance of such devices.
Emitter-nanoantenna coupling is typically accomplished by uncontrolled distribution of emitters into  nanophotonic structures.
Although potentially a route to realizing multiple emitter-antenna devices in a single step, the resulting  random statistical placement results in a low probability that a single emitter is associated with a bullseye, rather than zero or multiple emitters, and an even lower probability that such emitters are well centered within the bullseye. 
Specifically, we showed that insuﬃcient centering accuracy leads to imperfect collimation  and  thereby  collection  efficiencies  of  only  ~37\%  at  NA  =  0.65
\cite{Livneh2016,Harats2017DesignEmission}, which is far lower than the theoretical prediction of 85\%. \cite{Abudayyeh2017QuantumSources}. 

Alternatively, random distribution of emitters can be replaced by direct placement employing scanning probe techniques. For example, the pick-and-place technique allows a pre-characterized single emitter to be picked up from a surface and  moved  to  a  desired  location somewhere else on the surface with sub-$\mu$m precision. \cite{Dawood2018TheNanolithography,Schell2011ADevices,Nikolay2018AccurateStructures,Bohm2019On-chipPlatform,Zadeh2016DeterministicCircuits,Kim2017HybridChip}.
However, this technique   is not inherently scalable, as it relies on  locating  a  single  nanocrystal and identifying it as  emissive, adhering it  to  the  tip  of  an  atomic  force  microscopy  probe,  and  depositing  it  onto a device of interest.
Though a valuable technique for proof-of-principle experiments in the laboratory, this approach is not ultimately practical.

Here, we employ a different scanning probe technique, namely, dip-pen nanolithography (DPN). 
With DPN, we are able to directly "write" single nanocrystals with high precision into devices using liquid "inks". Because the QDs are maintained in their native solution-phase environment throughout the placement process, this integration step does not damage the nanocrystals (see Results). Importantly, in contrast with previously employed scanning-probe techniques, DPN is amenable to scaling, as multiple antennas can be targeted one after the other in a single writing step. 
Combined with detailed 3D FDTD optimization of the bullseye structures \cite{Abudayyeh2017QuantumSources} and a superior fabrication technique  \cite{Nagpal2009UltrasmoothMetamaterials.} as well as distinct improvements in the QD emitter itself,\cite{Chen2008GiantBlinking} we are able to show that an optimized coupling between emitter and antenna can, indeed, result in near-unity collection efficiencies.

Significantly, record collection efficiencies  of $>80\%$ into a low NA of 0.5 are paired with brightness levels of up to 0.76 photons/pulse (distributed between exciton and bi-exciton emission) and with high single photon purities achieved using temporal filtering.  
The results lay the groundwork for photonics-enabled quantum and communication technologies that rely on eﬃcient room-temperature nanoantenna- emitter devices.

\begin{figure*}[t]
  \centering
    \includegraphics{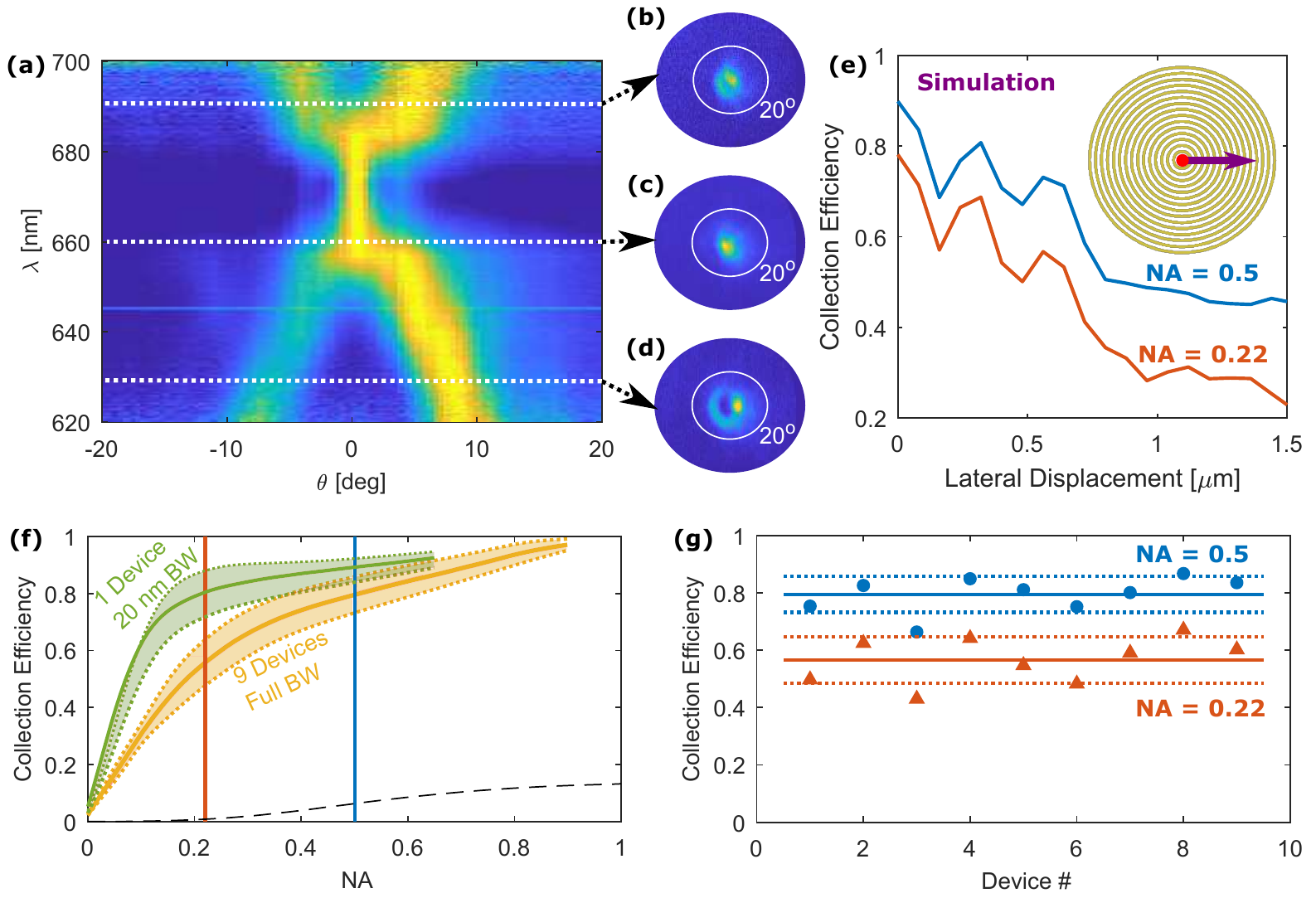}
    \caption{(a) Spectrally resolved back focal plane PL image of a final device. The image is normalized with respect to the spectrum, i.e. each line (wavelength) is normalized to the max in that line. (b),(c) and (d) show back focal plane images at $\lambda =$630, 660, and 690 nm respectively.(e) Simulated collection efficiency at various lateral displacements of a randomly oriented dipolar emitter from the center of the bullseye antenna for NA = 0.22, 0.5. (f) Green: collection efficiency of same device in (a)-(d) spectrally filtered in the spectral range between 660-680 nm. Yellow: average measured collection efficiency of 9 fabricated single emitter devices, each integrated over the whole QD PL spectral range. Black: FDTD simulation of a reference QD placed on a glass slide. (g) Measured collection efficiency for each of the 9 devices for NA= 0.22 (orange triangles), and 0.5 (blue dots). Solid lines represent mean values and dotted lines are 1 standard deviation from the mean.  \label{Fig: spectrallyresolved}}
\end{figure*}

\section{Results}
\subsection{Device design and fabrication}
We previously conducted detailed 3D FDTD simulations of the metal-dielectric bullseye antenna to optimize the device's directionality \cite{Abudayyeh2017QuantumSources}.
A recently reported method, known as template stripping, was used to fabricated the designed structures. \cite{Nagpal2009UltrasmoothMetamaterials.} 
Compared to our prior experimental work \cite{Livneh2016}, the resulting bullseye antennas were ultrasmooth. 
A first dielectric layer of Polymethyl methacrylate (PMMA) was then spin-coated in order to ensure that the QDs are placed at the correct height, followed by a thin layer of aluminum oxide for stability (see below).

Single QDs were then deposited onto the antennas following the sequential procedure outlined in \ref{fig: intro}. 
Briefly, a target region comprising multiple antennas is scanned with the AFM tip.
The resulting image constitutes a nanoscale topographical map that is used to guide the subsequent writing step. 
An AFM probe is then wetted with a QD-solvent suspension (the "ink"). 
In this work, we employ QDs that are known to be non-blinking and non-photobleaching at room temperature. These are CdSe/CdS core/thick-shell or "giant" QDs (gQDs) with an overall diameter of 17-19 nm (core is 5 nm). \cite{Chen2008GiantBlinking,Ghosh2012NewDots,Orfield2018PhotophysicsNanocrystals} 
The tip is wetted with a gQD-solvent suspension (“ink”) and a droplet is placed in the predetermined location by contacting the AFM tip to the substrate.
Significantly, this process does not damage or change the gQD optical properties (see Supplementary Information Figure S5).
Several variables of the DPN process affect the area of the deposition spot and the number of nanocrystals that are deposited per spot. 
These include dwell time, ink-substrate interactions, ink volume on the tip \cite{Dawood2018TheNanolithography} and the carrier liquid is the  high-boiling  solvent,  dichlorobenzene  (DCB).
Use  of  DCB allows the ink to remain a liquid throughout the writing process, such that fluid flow from tip to surface governs transport of QDs onto the substrate, with dwell time, ink-substrate interactions, and ink volume on the tip governing the volume of the deposited droplet.
Deposited-droplet   volume,   along   with   QD  concentration   in   the   ink,   then   determines the number  of  QDs transferred to the device.

Prior  to  writing   onto   devices,   the   influence   of   the   various   parameters   governing   liquid transport on the number of QDs placed in a writing step (e.g., $>$5 QDs, clusters of 2-3 QDs,  a  single  QD,  or  no  QD)  was  assessed  on  a  flat  (featureless)  region  of  the  substrate. 
In  this  case,  arrays   of droplet   spots   were   deposited   and   the   number   of    QDs    per   spot was  assessed.
In an  array  optimized  for  low  QD  delivery,  of 20 spots, 3 spots contained small clusters (2-3 gQDs) and 2 spots contained single gQDs (as determined by time-gated $g^{(2)}(0)$ \cite{Mangum2013DisentanglingExperiments.}), giving the percentages shown in Fig \ref{fig: intro}d. 
Tip “bleeding” (release of excess ink from the AFM tip and upper regions of the cantilever prior to writing), short contact times, and a low concentration of gQDs in the carrier solvent were combined to achieve this delivery of single gQDs and small clusters with a 10\% and 15\% success rate respectfully.\bibnote[supp]{See the Supplementary Material for more details}
We note that our demonstrated placement accuracy - about 300 nm \cite{supp}  - is an advance compared to pick-and-place approaches for which a placement accuracy of 500 nm has been reported. \cite{Zadeh2016DeterministicCircuits}

While both the precision and yield of placement is significantly improved compared to a random distribution approach \cite{Livneh2016}, there is a need to increase the percentage of single-QD devices created in a single writing step. To this end, we are integrating a dual scanning probe microscope with an optical microscope. A conventional AFM probe will enable high-resolution topography, while a nanopipette will deliver QD solutions, similar to how the DPN was implemented here. The in situ optical microscope will allow simultaneous determination of the success or failure of a writing step and, thereby, immediate correction by a second writing step as needed, all without changing probes or removing the substrate for analysis in a second optical system.

Interestingly, we found it necessary to modify the surface of the device to prevent the high-boiling DCB solvent from compramising the polymer dielectric layer. 
Specifically, initial successful attempts at placing single QDs into bullseye antennas were thwarted by rapid photobleaching of the emitters. 
It was observed that the polymer used to encapsulate the metal bullseye (serving as the 'lower half' of the dielectric waveguide layer)  was softened by the DPN solvent, causing the QDs to sink through the dielectric layer toward the metal layer over time, which resulted in quenching of QD emission. 
This was confirmed in control experiments that compared QD stability when deposited from DCB either on polymer-coated metal-on-glass or on polymer-coated glass. 
The nanocrystals placed on the polymer overcoating a Ag film quenched over time, while those placed on the polymer overcoating only glass did not quench. \cite{supp}
For this reason, we utilized a novel room-temperature pulsed gas-phase deposition technique process to deposit thin AlO$_x$ films\cite{Bose2018EngineeringApplications} onto the devices before conducting gQD deposition. \cite{supp}
This step did not damage the polymer and effectively “hardened” it to the DPN solvent, allowing sufficient stability to be maintained.

\subsection{Device Performance}
For devices in which there is no lifetime reduction, as is the case here, the collection efficiency at the first lens is what determines the overall brightness of the source.
Therefore, to characterize the overall performance of the emitter-antenna devices we conducted a series of directionality tests by imaging  photoluminescence (PL) of the QDs in the different devices at the back focal plane of our objective lens (NA=0.9) either on the slit of a spectrometer (with a CCD at the output) or directly onto a CMOS camera. 
In the first case the result will be a hyperspectral image of the back focal plane displaying the spectral dependence of the angular intensity distribution function which we will denote as $S(\lambda,\theta,\phi)$ where $\lambda$ is the wavelength and $\theta$ and $\phi$ are the polar and azimuthal angles respectively.
Fig \ref{Fig: spectrallyresolved}a is an image for $S(\lambda,\theta,\pi/2)$ of a typical device, which shows the broadband resonance of the antenna at between 660-680 nm. 
Moreover most of the QD emission over the entire 80 nm range shown in this image is within 10$^\circ$ from the normal emphasizing the highly broadband operation of these antennas.
This is highlighted in the spectral sections in Fig \ref{Fig: spectrallyresolved}b ($S(630 nm,\theta,\phi)$), Fig\ref{Fig: spectrallyresolved}c ($S(660 nm,\theta,\phi)$), and Fig \ref{Fig: spectrallyresolved}d ($S(690 nm,\theta,\phi)$).

\begin{figure*}[t]
  \centering
    \includegraphics{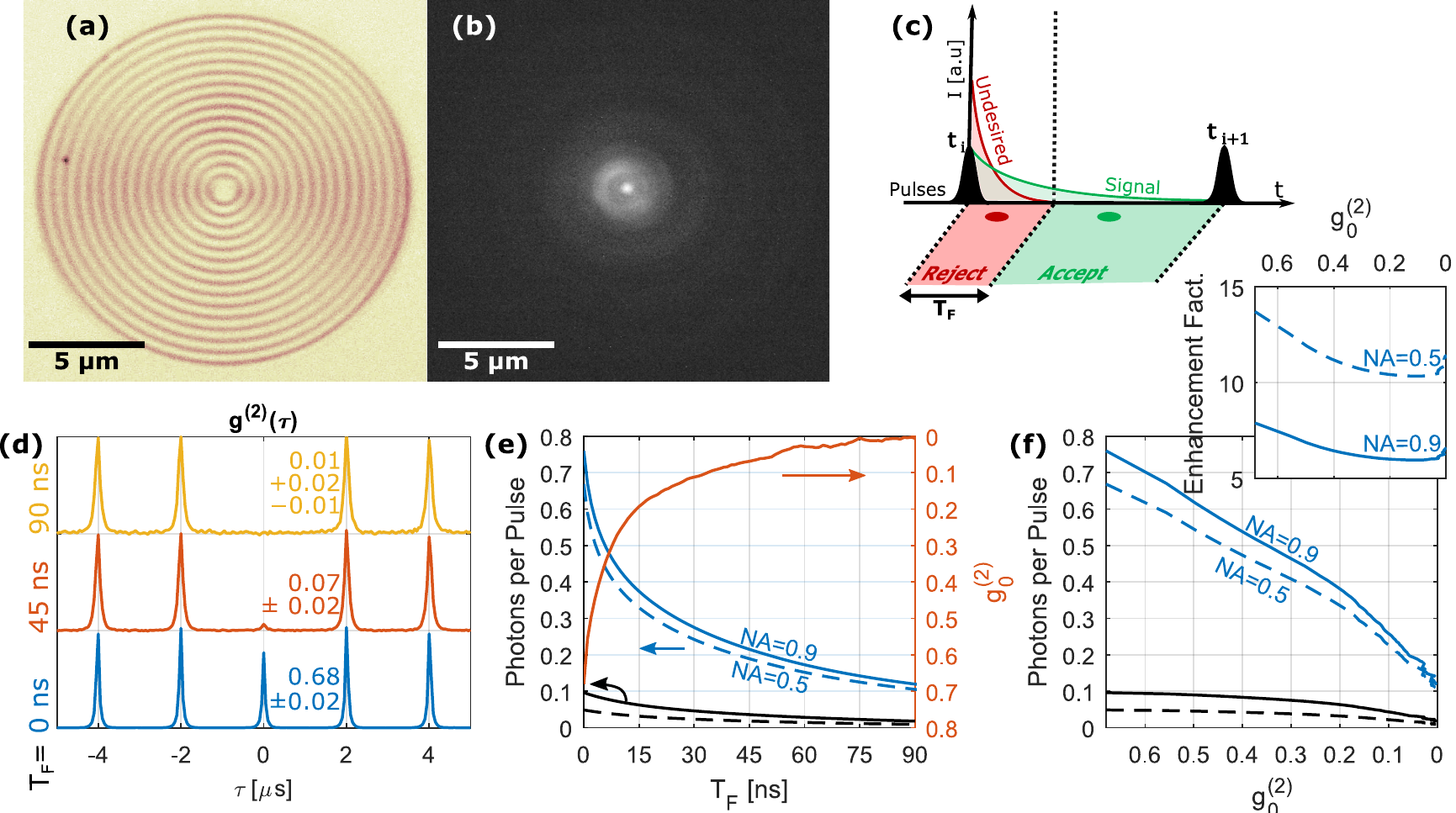}
    \caption{Results from a single QD coupled to a bullseye nanoantenna.(a) Microscope image and (b) Spatial PL map of a single photon device. 
    (c) Schematic displaying the time-gated filtering technique: the black pulses marks the laser excitation pulses, and the red and green lines mark the statistical temporal distribution of the bi-exciton and exciton emission with lifetimes of 11 ns and 67 ns respectively. Counts from $t=0$ to $t=T_F$ are post-rejected after the measurement. \cite{supp}
    (d) $g^{(2)}(\tau)$ for three values of $T_F$= 0,45, and 90 ns.
    (e) Brightness, and $g_0^{(2)}$ as a function of the time-gate delay ($T_F$) for a collection NA of 0.9 (solid lines) and 0.5 (dashed lines) compared to a similar QD placed on glass (black lines). (f) Brightness as a function of single photon purity. The inset represents the enhancement of the rate of a QD in our device compared to a similar QD on glass for an NA = 0.5 and 0.9. 
    \label{Fig: singleQD}}
\end{figure*}

Having shown the broadband directional nature of our antenna we move on to analyze the performance over the spectrally integrated angular intensity distribution $I(\theta,\phi)=\int S(\lambda,\theta,\phi) d\lambda $.
By integrating over all angles within a collection cone of a given NA one can calculate the ratio of photons collected; denoted as the collection efficiency:
\begin{equation}
    \eta = \frac{\int_0^{2\pi}d\phi\int_0^{\theta_{NA}}  d\theta \sin(\theta) I(\theta,\phi)}{\int_0^{2\pi}d\phi\int_0^{\frac{\pi}{2}}d\theta \sin(\theta) I(\theta,\phi)}
\end{equation}
To demonstrate the reproducibility of our method we display in Fig. \ref{Fig: spectrallyresolved}f \& g  the measured collection efficiency for 9 different emitter-antenna devices (4 of which contained single QDs) which were made under identical conditions. 
These results are compared to the reference collection efficiency of an emitter in free space.
The results show significant directionality enhancement and good reproducibility among the devices.
The two NAs (0.22,0.5) shown in Fig. \ref{Fig: spectrallyresolved}g represent the NAs of commercially available fibers indicating the promise of these devices for direct coupling to optical fibers.
Furthermore to emphasize the importance of our high precision positioning on the performance of the devices we compare in Fig. \ref{Fig: spectrallyresolved}e the collection efficiency of our devices with FDTD simulations \cite{LumericalInc.} of a randomly oriented emitter displaced laterally from the center. 
A significant degradation of performance is observed for even sub-500 nm displacements, with sub-micron displacements being devastating to performance, thus demonstrating the need for such new integration methods as described here.

Fig. \ref{Fig: singleQD} displays the results for an antenna with a single QD placed in the center. 
In Fig. \ref{Fig: singleQD}a,b the spatial images of the device and of the PL distribution are distinguished. 
We measure photon brightness levels of up to 0.76 photons/pulse. This corresponds to a photon rate of 380 kHz (330 kHz) at an optical pump rate of 500 kHz or a maximal photon rate of  $\sim$ 11.5 MHz (10.1 MHz) at the CW limit for a collection NA of 0.9 (0.5). Accordingly the bi-exciton and exciton quantum yields (lifetimes) are 21\% (11 ns) and 55\% (67 ns) respectively. \cite{supp}
With no temporal filtering the value of $g^{(2)}(0) = 0.68$.
This high value of $g^{(2)}(0)$ is a result of the measurements being mainly done near saturation where both bi-exciton and plasmonic emission play a role. \cite{supp}
Using the time-filtered gating technique \cite{Mangum2013DisentanglingExperiments.,Feng2017PurificationDots,Abudayyeh2019PurificationSources} we confirm that the emitter is indeed a single QD by measuring the second order coherence starting from various times ($T_F$) after the excitation pulse.
This helps in differentiating between the emission of bi-excitons (which have significantly shorter lifetimes) and the emission of multiple QDs \cite{Mangum2013DisentanglingExperiments.}. 
Using this technique the value of $g^{(2)}(0)$ quickly drops below 0.5 after only 2 ns of temporal filtering and continues falling as $T_F$ increases resulting in high single photon purity for larger filtering times reaching $g^{(2)}(0)<0.01$ with 0.12 photons per pulse at $T_F=90$ ns .
Although this increase in single photon purity comes at the cost of brightness, one might tune this brightness-purity trade-off to optimize the specific application of interest as shown in Fig.\ref{Fig: singleQD}f.
Furthermore we recently proposed a new heralding scheme to overcome this limitation \cite{Abudayyeh2019PurificationSources}  which  utilizes  the  difference  in  lifetime between  the  bi-exciton  and  exciton  states  and/or  the  cascaded  nature  of  the  emission  to separate the resulting two photon emission in time.  Critically we have shown that the use of these methods can allow record high single photon purities with much lower loss of efficiency or rate.

Finally, we emphasize the benefit of our device over a similar QD on glass.
We compare in Fig. \ref{Fig: singleQD}e,f between the rate of a QD in our device and on glass. We also display this enhancement of the rate for different $g^{(2)}(0)$ in the inset of Fig. \ref{Fig: singleQD}f where a factor of 10 improvement in rate is displayed over a QD on glass \textit{even for the highest single photon purities}. 

\begin{figure}[t]
  \centering
    \includegraphics{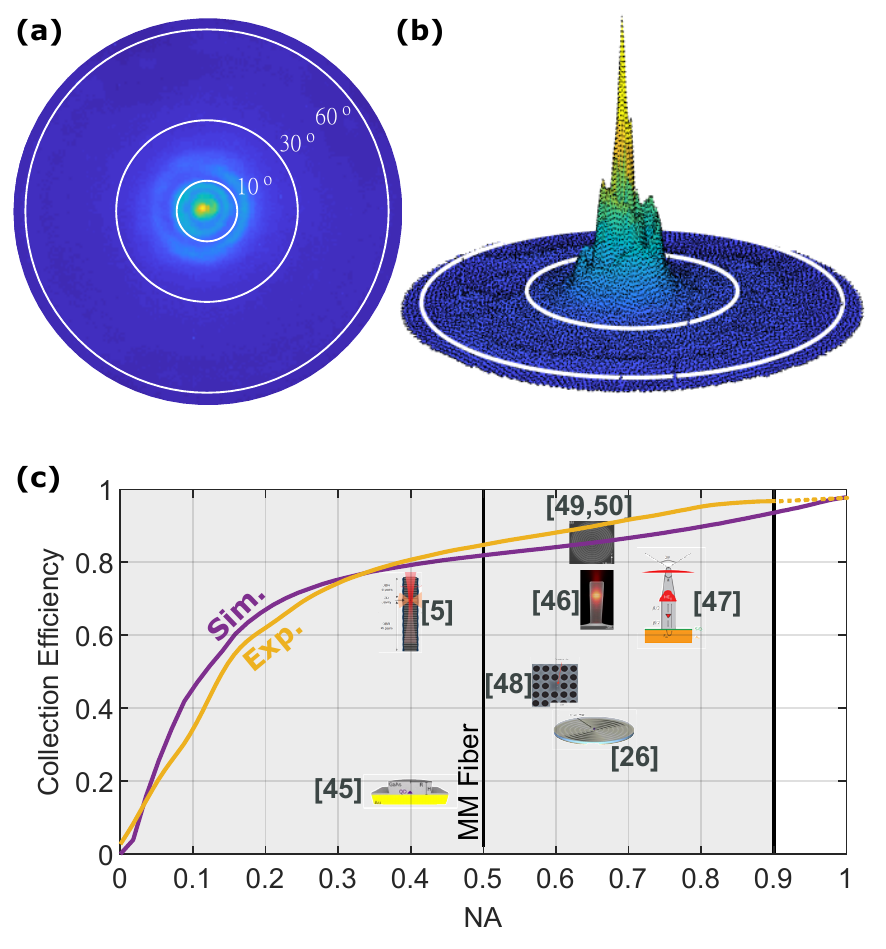}
    \caption{Results from a single QD coupled to a bullseye nanoantenna. (a) and (b)  Back focal plane image of single photon device. (c) Collection efficiency of device compared to simulation measurements and various state-of-the-art platforms: Micropillar cavities \cite{Fischbach2017EfficientMirror,Gazzano2013BrightPhotons,Liu2017AMethod}; Tapered Nanowire \cite{Claudon2010ANanowire};  Photonic crystal cavities \cite{Madsen2014EfficientCavity};  Bullseye nanoantenna \cite{Livneh2016,Liu2019AIndistinguishability,Wang2019On-DemandIndistinguishability}.
    \label{Fig: singleQD_kspace}}
\end{figure}

The reasons behind this high relative photon flux even for low NAs can be determined by inspecting the back focal plane image from such a device. As can be seen in Fig. \ref{Fig: singleQD_kspace}a,b, the emission displays strong directional behaviour as most of the emission is to low angles. 
This results in very high collection efficiencies into low NA (Fig. \ref{Fig: singleQD_kspace}c). 
We compare our results to detailed FDTD simulations \cite{LumericalInc.} taking into account the broadband nature and random dipole orientation of the QDs and find good agreement between simulation and experiments, as is shown in Fig. \ref{Fig: singleQD_kspace}c (See Ref. \cite{Abudayyeh2017QuantumSources} for further details on the simulation).

\section{Discussion}
The described approach enables performance enhancements compared to previously reported state-of-the-art emitter-antenna platforms, including low-temperature platforms (Fig. \ref{Fig: singleQD_kspace}c). The collection eﬃciencies we observe are higher than those achieved by micropillar cavities \cite{Fischbach2017EfficientMirror,Gazzano2013BrightPhotons,Liu2017AMethod}, tapered nanowires \cite{Claudon2010ANanowire},  photonic crystal cavities \cite{Madsen2014EfficientCavity}, and bullseye antennas \cite{Liu2019AIndistinguishability,Wang2019On-DemandIndistinguishability}.
Notably, most of these  devices  operate  at  cryogenic  temperatures with  cryogenic  sources,  where  the emission lines are much narrower allowing for easier photon collection. 
In contrast our sources (and most room-temperature sources) have a  broader spectrum (30-40 nm FWHM\cite{supp}). 
Despite this, the optimized design of the bullseye antenna combined with the ability to accurately target the bullseye center enables record high photon collection efficiency over a broad spectral range as shown in Fig. \ref{Fig: spectrallyresolved} a-e. 
This is a key achievement  that will be applicable to all room-temperature sources.  

In addition, the approach shown here is especially suited for collection into lower NAs , such as is needed for direct coupling into optical ﬁbers or for free space transmission to remote nodes.
For example, the number of collected photons into an NA of 0.12 (0.22), corresponding to the NA of a single (multi) mode fiber, is enhanced by a factor of 6 (5) compared to an emitter in free space. 
This is also highlighted in Fig. \ref{Fig: singleQD}a in the small difference in the collected brightness into an NA of 0.5 and 0.9.

In comparing the current work with a previous work from our group on bullseye nanoantennas,\cite{Livneh2016} we note that the earlier work could not show very high collection efficiencies. 
Namely, a maximum of 37\% collection eﬃciency was realized at an NA of 0.65, compared to a 90\% collection eﬃciency into the same NA here.
This significant improvement was achieved due to multiple improvements including: (a) detailed optimization of the structure based on 3D FDTD simulations \cite{Abudayyeh2017QuantumSources}; (b) a new template stripping method used for the fabrication on the metal part of the antennas,  yielding a much higher quality antennas \cite{supp}; (c) the novel placement method described above yielding a much better emitter positioning; and (d) a higher signal to noise ratio and better loss estimation methods afforded by the combined use of brighter and non-blinking QDs \cite{Orfield2018PhotophysicsNanocrystals} and a higher NA objective. 

In summary we have put forward a method for placing single QDs onto predefined substrates with high spatial accuracy. 
By using this method with metal-dielectric hybrid bullseye antennas we were able to reproducibly show record-high directionalities of single photons at room temperature compared to other state-of-the-art platforms.
Collection efficiencies exceeded 80\% for NA=0.5 and brightness levels were up to 0.76 photons/pulse (distributed between exciton and bi-exciton emission), with high single photon purities achieved using temporal post-selection. 
Thus, a viable path forward for both ﬁber-based and free space quantum communication applications operating at room-temperature is demonstrated.

\section*{Supplementary Material:}
See supplementary material for details on (1) synthesis and fabrication; and (2) experimental details including the measurement scheme, photon rate calculation, and collection efficiency calculation. 

\section*{Acknowledgements}
This work was performed in part at the Center for Integrated Nanotechnologies (CINT), a Nanoscale
Science Research Center and User Facility operated for the U.S. Department of Energy (DOE) Office of
Science. AB and JW were funded by CINT. All other Los Alamos National Laboratory (LANL) authors were
funded by the Laboratory Directed Research and Development (LDRD) program, grant 20170001DR. LANL,
an affirmative action equal opportunity employer, is operated by Los Alamos National Security, LLC, for the
National Nuclear Security Administration of the U.S. Department of Energy under contract DEAC52-06NA25396.
R. B.. and A.V.M. were supported by U.S. Department of Energy, Office of Basic Energy Sciences, Division of Materials Sciences and Engineering under Award No. DE-SC0010697

\section*{Data Availability}
The data that support the findings of this study are available from the corresponding author
upon reasonable request.
\bibliography{mendeley}
\end{document}


\title{SUPPLEMENTARY MATERIAL: \\  
Single Photon Sources with Near Unity Collection Efficiencies by Deterministic Placement of Quantum Dots in Nanoantennas
}

\author{Hamza Abudayyeh}
\author{Boaz Lubotzky}
\affiliation{Racah Institute of Physics, The Hebrew University of Jerusalem, Jerusalem 9190401, Israel}
\affiliation{The Center for Nanoscience and Nanotechnology, The Hebrew University of Jerusalem, Jerusalem 9190401, Israel}

\author{Anastasia Blake}
\author{Jun Wang}
\author{Somak Majumder}
\author{Zhongjian Hu}
\author{Younghee Kim}
\author{Han Htoon}
\affiliation{Materials Physics \& Applications Division: Center for Integrated Nanotechnologies, Los Alamos National Laboratory, Los Alamos, New Mexico 87545, United States}

\author{Riya Bose}
\author{Anton V. Malko}
\affiliation{Department of Physics, University of Texas at Dallas, Richardson, Texas 75080, United States}

\author{Jennifer A. Hollingsworth}
\email{jenn@lanl.gov}
\affiliation{Materials Physics \& Applications Division: Center for Integrated Nanotechnologies, Los Alamos National Laboratory, Los Alamos, New Mexico 87545, United States}

\author{Ronen Rapaport}
\email{ronenr@phys.huji.ac.il}
\affiliation{Racah Institute of Physics, The Hebrew University of Jerusalem, Jerusalem 9190401, Israel}
\affiliation{The Center for Nanoscience and Nanotechnology, The Hebrew University of Jerusalem, Jerusalem 9190401, Israel}
\affiliation{The Applied Physics Department, The Hebrew University of Jerusalem, Jerusalem 9190401, Israel}

\date{\today}

\maketitle

\section*{Synthesis and Fabrication}
\begin{figure}[t]
  \centering
    \includegraphics{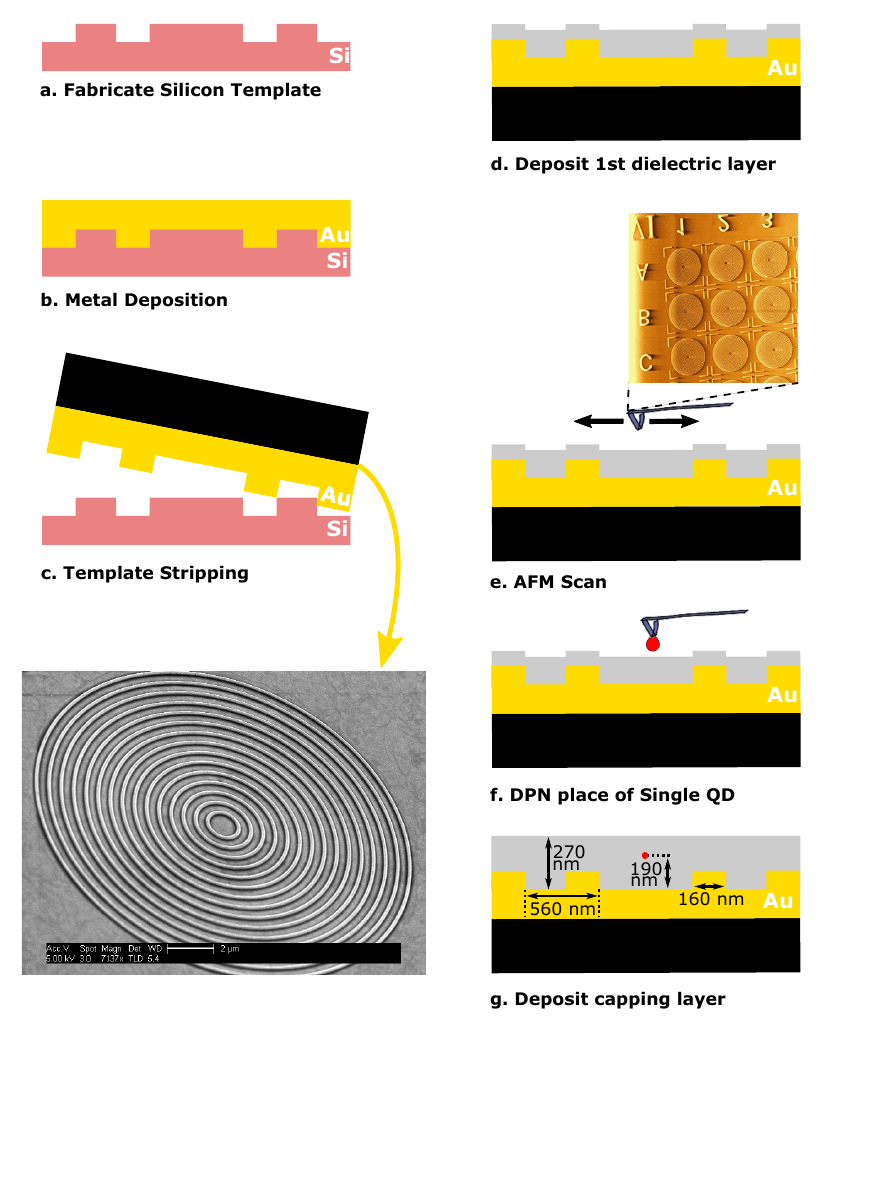}
    \caption{Step by step fabrication process
    \label{Fig: fabrication}}
\end{figure}
\subsection*{Synthesis of emitters}
\subsubsection*{Materials}
Cadmium oxide (CdO, 99.99\%), oleic acid (OAc, 90\%), 1-octanethiol (OT, 98\%) and selenium
(1-3 mm shot, 99.99 \%) were purchased from Alfa Aesar. 1-octadecene (ODE, 90\%) were purchased from Acros Organics. Trioctylphosphine (TOP, 97\%) trioctylphosphine oxide (TOPO, 99\%) and oleylamine (OAm, 70\%) were purchased from Sigma Aldrich. Octadecylphosphonic acid (ODPA, 97\%) was purchased from Strem. All chemicals were used as received without further purification.

\subsubsection*{Synthesis of CdSe QDs}
CdSe quantum dot (QD) cores were synthesized as follows. CdO (60 mg), TOPO (3 g) and ODPA(280 mg) were combined and degassed (under vacuum) for 1 h at \SI{125}{\degreeCelsius}. 
Following degassing, the reaction mixture was heated to \SI{330}{\degreeCelsius} under Ar atmosphere, when the color of the solution was found to change from reddish brown to clear and colorless indicating formation of Cd-ODPA complex. At this point, 1 mL of TOP was injected and the reaction mixture was
heated to \SI{380}{\degreeCelsius}. When the temperature stabilized, Se (60 mg) dissolved in TOP (0.5 mL)
(prepared separately in an inert atmosphere glove box) was rapidly injected into the reaction mixture to induce nucleation of CdSe QDs. After 1.5 min of growth, the heating mantle
was removed and the reaction was allowed to cool to room temperature. To prevent
solidification of TOPO, approximately 2 mL hexane was added when the temperature reached about \SI{60}{\degreeCelsius}. 
The QDs were then purified of excess ligands and unreacted precursors by multiple acetone precipitation/hexane redissolution cycles. The synthesis yielded cores with first absorption maxima $>$600 nm. Final QDs were dried and dispersed in hexane.

\subsubsection*{Synthesis of giant CdSe/CdS core/shell QDs (gQDs)}
CdSe/CdS gQDs (core diameter 5 nm, overall diameter 17-19 nm) were synthesized via a modified continuous injection process.\cite{Chen2013CompactBlinking} In a typical synthesis, $1.0 \times 10^{-7}$ mol of CdSe QDs were mixed with 3 mL of OAm and 3 mL of ODE and degassed for 0.5 h at room temperature and 1 h at \SI{80}{\degreeCelsius}.
Following degassing, the reaction mixture was heated to \SI{310}{\degreeCelsius} (\SI{15}{\degreeCelsius}/min) under Ar, during the course of which Cd-oleate (0.1 M, 5 mL) and OT (0.12 M in ODE, 5 mL) were simultaneously injected at the rate of 3 mL/h starting at \SI{240}{\degreeCelsius}. 
On completion of shell precursor addition, 1 mL of OAc was swiftly injected and the reaction mixture held at \SI{310}{\degreeCelsius} for 45-60 min. The reaction was then allowed to cool to \SI{90}{\degreeCelsius} for degassing. When the temperature reached \SI{105}{\degreeCelsius}, the vessel was placed under dynamic vacuum for 30 min, or until bubbling ceased. 
Multiple injections, following the above process, of 5 mL portions of Cd oleate and OT precursors were performed to yield the desired CdS shell thickness. 
The synthesized QDs were precipitated from growth solution by ethanol and redissolved in hexane. 

\subsection*{Device Fabrication}
The fabrication  process is detailed in Fig. \ref{Fig: fabrication}. 
The antenna's parameters were optimized in our previous theoretical study \cite{Abudayyeh2017QuantumSources}. The same parameters were used here and are shown in Fig. \ref{Fig: fabrication}g.

The metallic portion of the nanoantenna was fabricated using the template stripping method \cite{Nagpal2009UltrasmoothMetamaterials.}. 
The template was made of silicon using a Ga focused ion beam machine;
250 nm of gold was then deposited on the template, followed by spin coating of SU8 3010 at 3000 RPM which was prebaked at 95 C for 5 mins.
A glass slide was then attached and the SU8 was cured with UV at  150 mJ/cm$^2$ for 15 sec flood exposure.
The Au attached to the glass was stripped off the template resulting in highly smooth bullseye antennas (see the inset in Fig. \ref{Fig: fabrication}c).
The sample was then covered with 174 nm of PMMA followed by 5 nm ALD deposition of Al$_2$O$_3$.
The target areas were then scanned resulting in a topographical map (see the inset in Fig. \ref{Fig: fabrication}e). The tip was then inked and a droplet is placed at the center of the desired structure (see below for more details).
The QDs were protected by another 5 nm ALD deposition of AlO$_x$ and finally a capping layer of 70 nm of SiO$_2$ was then deposited on the sample.

\subsubsection*{Aligning DPN Substrates for Printing}
\begin{figure}
  \centering
    \includegraphics[width = \textwidth]{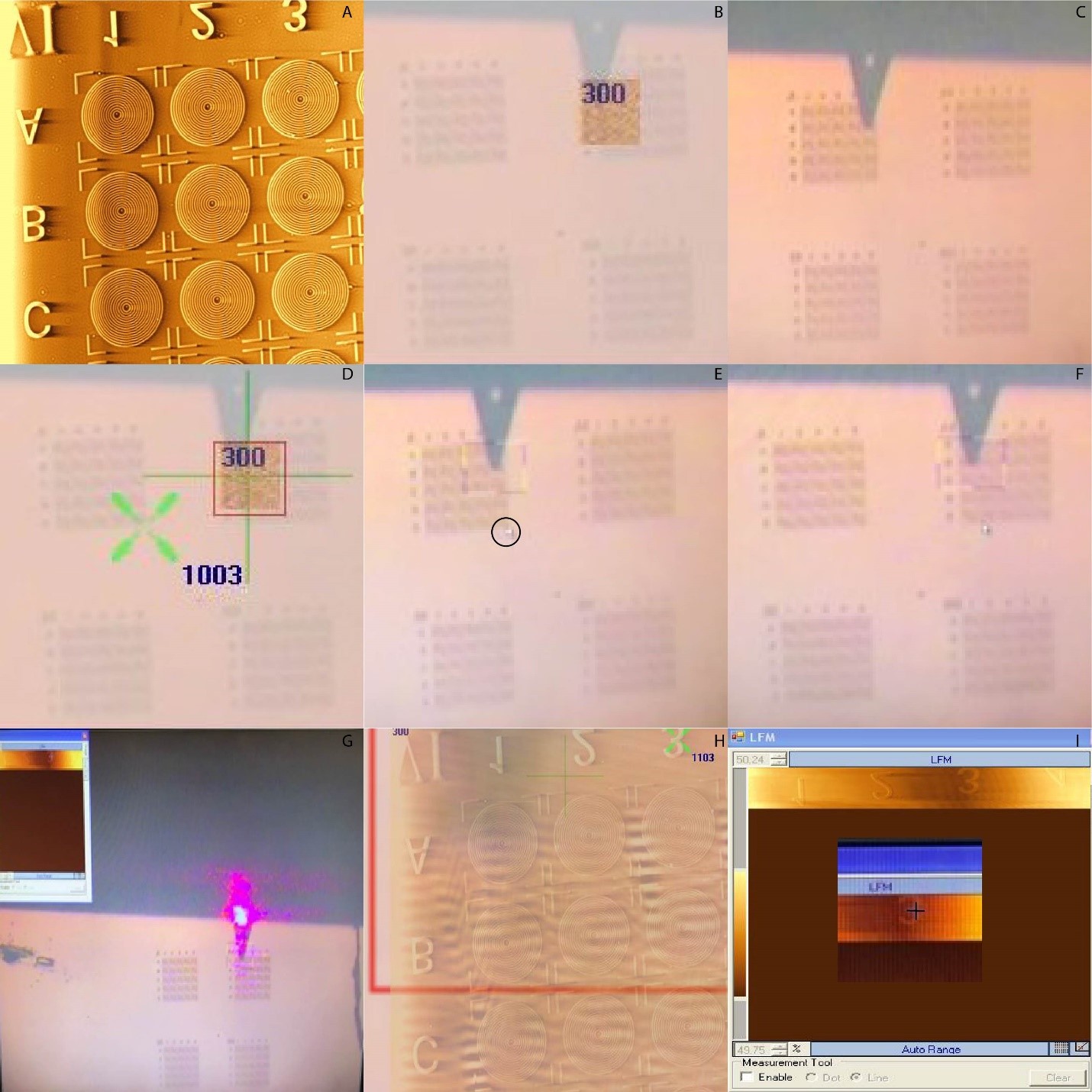}
    \caption{The DPN alignment process. A) AFM scanned image of the target area. B) Optical image of the scanned target area and surrounding substrate. C) Position of the tip after returning to the saved x, y, and z coordinates. D) Initial (optical) alignment step; selecting a feature visible in the original saved optical image. E) Initial (optical) alignment step; selecting the same feature selected in the optical image (D) from the live feed (cursor, black circle). F) Resulting position of the tip after initial/optical alignment. G) Re-scanning the target area and the resulting image (inset). H) Secondary (AFM) alignment step; selecting a visible feature in the original AFM scanned image (green x). I) Secondary (AFM) alignment step; selecting the same visible feature in the new scan of the target area (black cross). The tip position will be adjusted accordingly.
    \label{Fig: supp1}}
\end{figure}
\begin{figure}
  \centering
    \includegraphics[width = 0.5\textwidth]{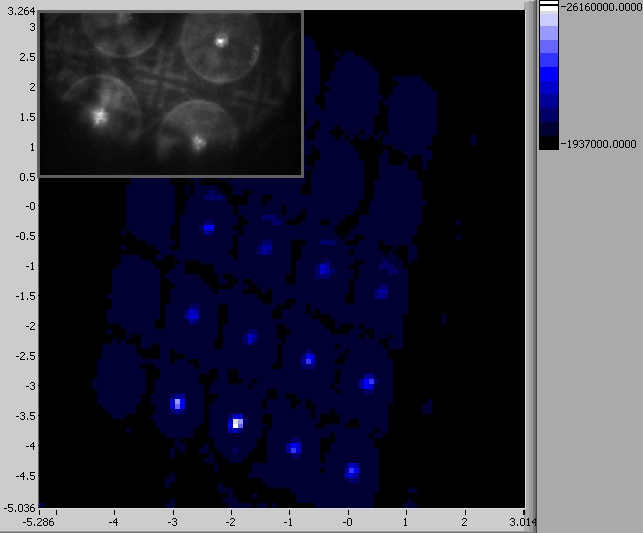}
    \caption{Photoluminescence image shows 12 antennas onto which gQDs were placed in the centers of each bullseye. Here, gQDs were present as small clusters, rather than single nanocrystals. Color scale represents emission intensity in counts, while the distance scale can be approximated by comparing to the overall antenna size of 16 mm. Inset shows a a spectral PL image.
    \label{Fig: confocalSI}}
\end{figure}
The target area for a print job must be scanned in its entirety before the tip is inked to provide an image on which a printing pattern can be drawn. 
Therefore, in order to be able to return to the same target area after inking, an alignment must be performed via the InkFinder program. 
After the target area is scanned and the image is saved (Fig. \ref{Fig: supp1} A), the system is taken out of feedback (the laser is turned off and the tip is retracted \SI{50}{\micro\metre} from the surface), a second image (optical) is saved from the retracted position, and the x, y, and z coordinates are also saved (Fig. \ref{Fig: supp1} B). 
Inking and bleeding is then performed, after which, the alignment process can begin. 
The tip is first returned to the correct x, y, and z positions. Despite having returned to the stored coordinate values, it can sometimes be seen optically that the positioning of the tip is not the same as it was at the end of the initial scan (Fig. \ref{Fig: supp1} C). 
The next step of the alignment process is to select a feature on the substrate visible in both the initial saved optical image from the retracted position of the tip and in the live feed (Fig. \ref{Fig: supp1} D \& E). 
InkFinder will then adjust the position of the tip based on this information (Fig. \ref{Fig: supp1} F).
This step in the alignment process is limited by how well focused and visible features on the substrate are in the optical images.
The final step in the alignment process begins by putting the system back in feedback (turning on the laser and approaching the surface) and re-scanning the target area of the substrate. 
Because the tip is now inked, the whole target area should not be scanned. 
Rather, the substrate should be initially oriented such that identifying markers can be scanned without going over the entire target area again. 
Once an identifying marker is seen on the new scan, the scanning process is stopped (Fig. \ref{Fig: supp1} G). 
InkFinder will then allow the user to select a feature visible in both the original scanned image and in the new scan, and adjust the position of the tip accordingly (Fig. \ref{Fig: supp1} H \& I). 
Limiting factors are that alignment features can only be clicked on once in each image, and only one feature can be used. 
If the user does not click on the exact same spot on the feature in each image, alignment will be affected accordingly.

\section*{Experimental Details}
\subsection*{Measurement Scheme}
\begin{figure}
  \centering
    \includegraphics{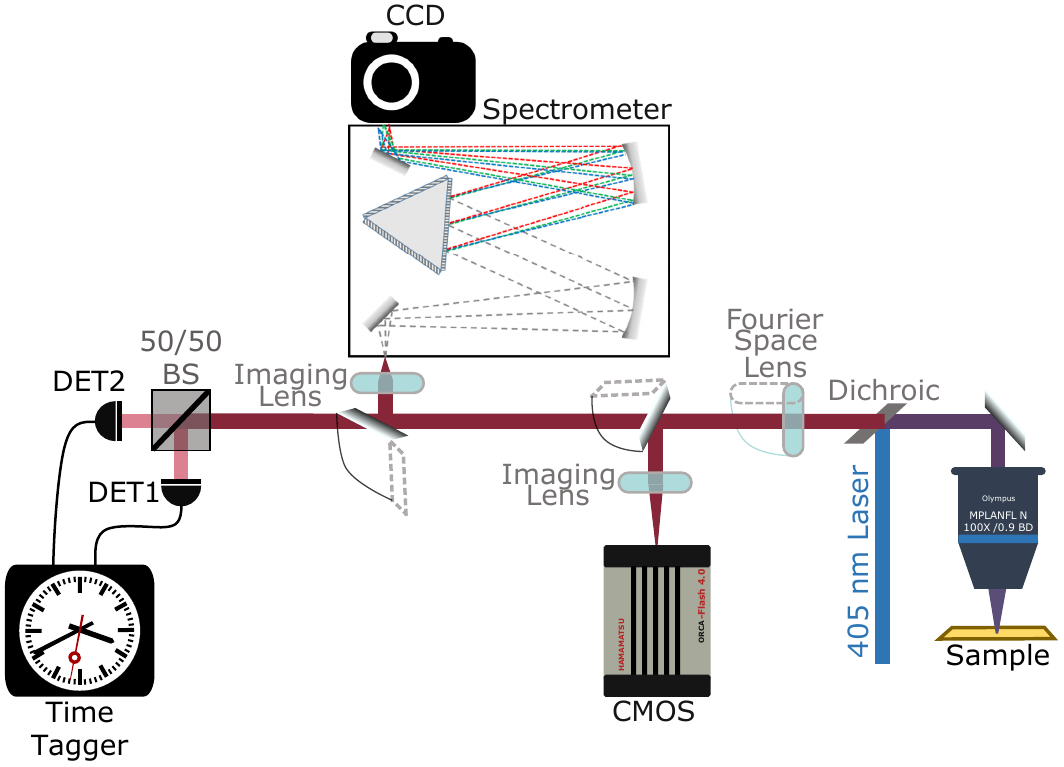}
    \caption{Diagram of experimental setup
    \label{Fig: setup}}
\end{figure}
The setup used for the PL measurements is shown in Fig. \ref{Fig: setup}.
The input laser light is a 0.5 - 2 MHz 405 nm pulsed
laser generated by second harmonic generation of a femtosecond Ti:Sapphire
laser operating at 810 nm. The sample is scanned using a
system of scanning stages and the laser light is focused onto the
sample using a 0.9 NA objective (Olympus MPLFLN100xBD). The emission 
is collected using the same objective and directed to the collection
arm using a 567 nm longppass dichroic mirror. The sample is then imaged
using a CMOS camera to verify the location with respect to the sample. 

For back-focal plane images the fourier-space lens is flipped-on and the image is taken either directly to a CMOS camera (for spectrally integrated back-focal plane images) or through a spectrometer to a CCD camera for spectrally resolved back-focal plane images. 

To conduct time resolved single photon counting measurements the emission is redirected to our Hanbury Brown Twiss module consisting of  a  beam-splitter and a set of single
photon avalanche photodiodes (Excelitas SPCM-AQRH-14-FC).
The signal from each detector is routed to a different channel on
the timetagger device (Swabian TimeTagger 2.0). The output from the timetagger is a vector containing the arrival times of all the photons within
the set exposure time with respect to the beginning of the measurement
in addition to a tag labeling which channel this count came from.
These arrival times are commonly referred to as global times. Another
channel on the time-tagger records the excitation pulse times. By
comparing the global time of each count with the nearest preceding
laser pulse one can find the local time for the count. A histogram
of these local times is what constitutes a lifetime measurement. Therefore
at the end of this step we have information about which pulse and
what channel the count came from, and its global and local times.
Using this information one may choose to conduct the correlation by post-selecting only  detector counts that arrive after a given period from the laser pulse. This is what is referred to as a Time-Gated filtering measurement in the main text. 

\subsection*{Spatial Accuracy Determination}
\begin{figure}
  \centering
    \includegraphics[width = \textwidth]{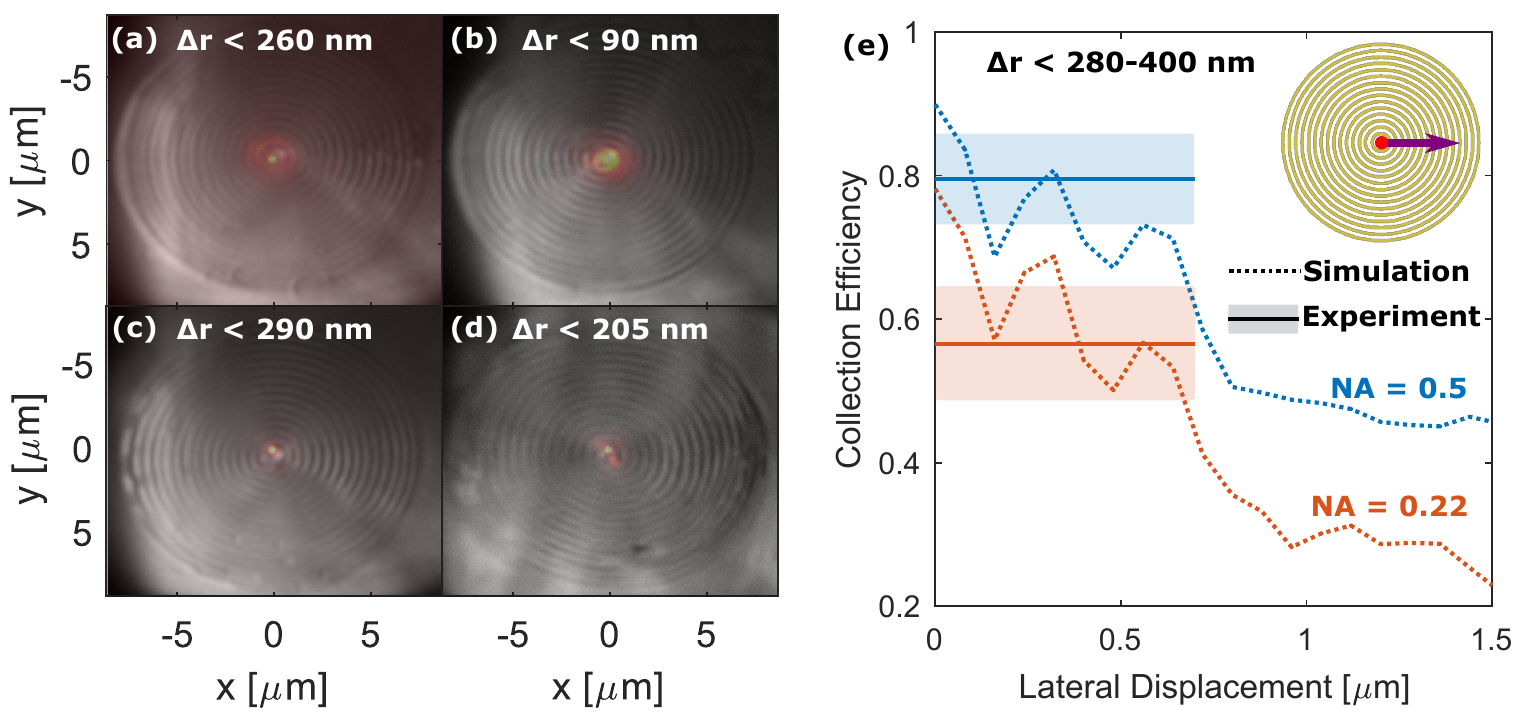}
    \caption{ (a-d) An overlay of a white-light microscope image showing the antenna and a QD fluorescence image (red) displaying the accuracy of spatial placement. (e) A comparison between simulations (dotted lines) and experimental data (solid lines) for estimating the lateral displacement of the emitters. The shaded areas in the experimental data represents one standard deviation from the mean.
    \label{Fig: spatialaccuracy}}
\end{figure}
The spatial accuracy was determined experimentally by taking near-simultaneous exposures of the antenna using either white light (to image the surface) or the excitation laser (to observe the fluorescence). 
By overlaying these two images, one on top of the other, we are able to determine an upper bound on the displacement between the QD emission and the center of the bullseye. 
This data is shown in Fig. \ref{Fig: spatialaccuracy}(a-d) from 4 different devices suggesting that the spatial accuracy is within a 300 nm radius. 
We should note that this accuracy is only an upper limit since there are some random drifts between subsequent exposures due to the mechanical action required to switch between white-light imaging and fluorescence imaging. 

This experimental data is in line with what we expect from simulations. 
In Fig. \ref{Fig: spatialaccuracy}e we plot the dependency of the collection efficiency on lateral displacements (for NA = 0.22 and 0.55) which is just Fig. 2e in the main text. 
On top of that we plot the measured collection efficiency for the same NAs (Fig.2g in the main text). 
This suggests that if the discrepancy between measured and simulated results was due only to lateral displacements, then we expect that on average the spatial accuracy would lie in the range of 280-400 nm. 
Of course the discrepancy can be due to any number of fabrication and experimental factors so this range remains an upper limit as well.

\subsection*{Photon rate calculation}
\begin{figure}
  \centering
    \includegraphics[width = \textwidth]{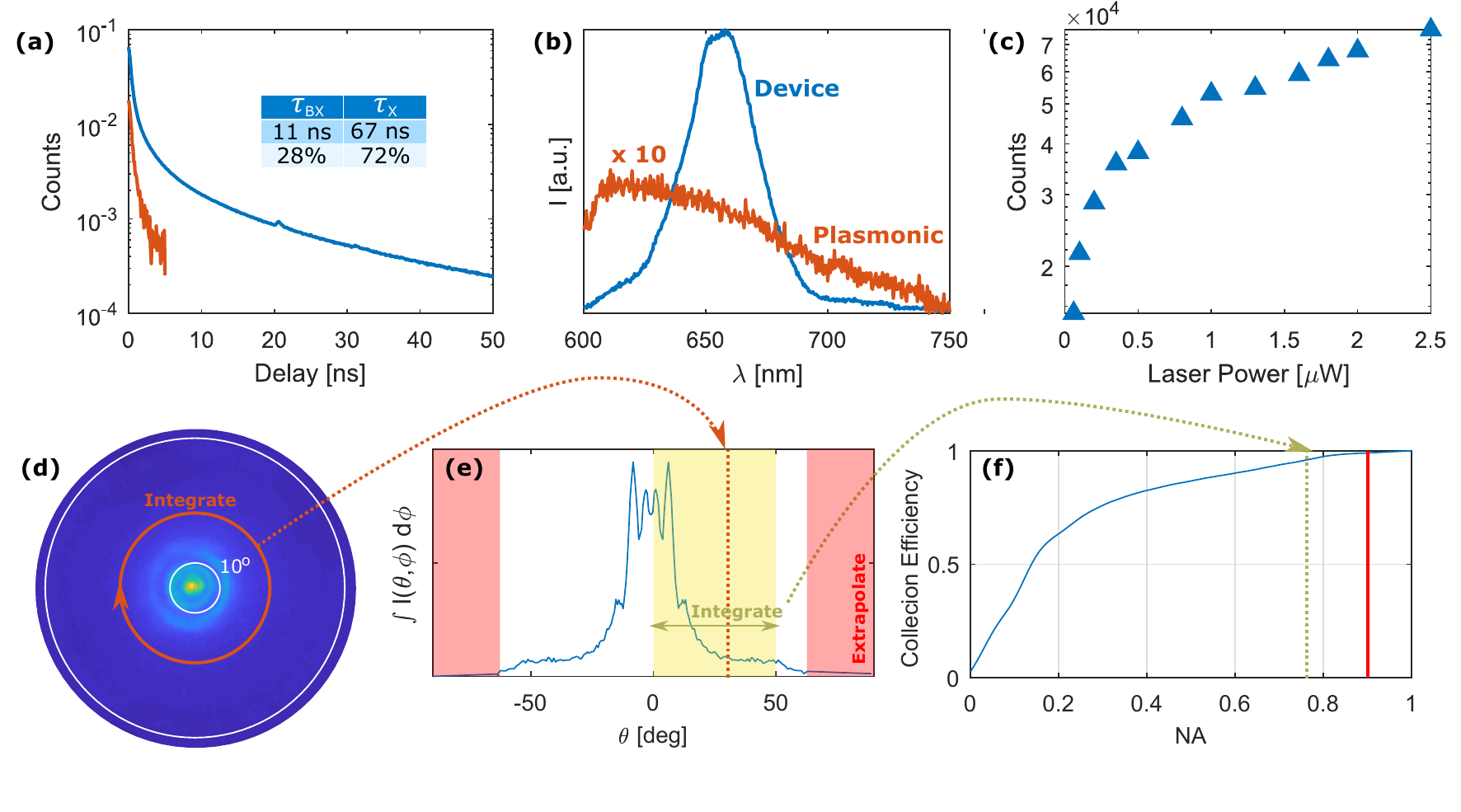}
    \caption{ (a) Lifetime, (b) spectrum, and (c) power saturation curves of an emitter-antenna device and an empty antenna displaying the plasmonic contribution. (d) Back-focal plane image, (e) azimuthally integrated angular intensity distribution, and (f) collection efficiency for emitter-antenna device.
    \label{Fig: estimation}}
\end{figure}

Part of the detected photon rate at the single photon detectors is due to plasmonic emission which reduces the purity of our source. 
To verify that this is indeed plasmonic emission rather than higher multiexcitons (other than biexcitons) we measure an antenna that was intentionally left empty and find the lifetime and spectrum shown in Fig. \ref{Fig: estimation}a and b. 
At the same excitation powers we find that the shortest component in our device's lifetime curve corresponds to the same lifetime and rate as in an empty antenna thus verifying that it is indeed due to plasmonic emission \cite{Livneh2016}. 
This is not spectrally filtered by our system due to the spectral overlap between the QD emission and the plasmonic emission. 
After confirming this we deduct this plasmonic rate from our measured photon rate above (Fig. \ref{Fig: estimation}b).
This emitter rate is then divided by the system efficiency (22.6\%; Table \ref{tab: Coll_eff}) to yield the photon rate into the NA of our objective. 
Dividing by the collection efficiency into our NA (which is nearly one see the main text and below) we reach to the photon rate emitted by the emitter in our device (0.76 photons/pulse).
To estimate the biexciton and exciton quantum yields we fit the lifetime curve of our emitter (after deducting plasmonic noise) to a biexponential function and find that 28\% of the emission is due to a shorter lifetime component (which is attributed to the bi-exciton) and 72\% is due to the exciton emission.
Therefore by taking into account the above photon rate the bi-exciton and exciton quantum yields are reported to be QY$_{BX}$ =0.21 and QY$_X$=0.55 respectively.
\begin{table}

\caption{System efficiency estimation}\label{tab: Coll_eff}
\begin{centering}
\begin{tabular}{|c|c|c|}
\hline 
Component & Method & Efficiency\tabularnewline
\hline 
\hline 
Objective transmission & meas & 90.0 \%\tabularnewline
\hline 
600 nm SP dichroic & meas & 96.6 \%\tabularnewline
\hline 
700 nm SP filter & meas & 88.5 \%\tabularnewline
\hline 
550 nm LP filter & meas & 95.2 \%\tabularnewline
\hline 
600 nm LP filter & meas & 84.3 \%\tabularnewline
\hline 
Beam splitters (2) & meas & 86.3 \%\tabularnewline
\hline 
Mirrors (6) & meas & 75.9 \%\tabularnewline
\hline 
Fiber coupling & meas & 80.0 \%\tabularnewline
\hline 
Detector efficiency & fact & 70.0 \%\tabularnewline
\hline 
\multicolumn{2}{|r|}{\textbf{Total}} & \textbf{22.6 \%}\tabularnewline
\hline 
\end{tabular}
\par\end{centering}

\end{table}

\begin{figure}
  \centering
    \includegraphics[width = \textwidth]{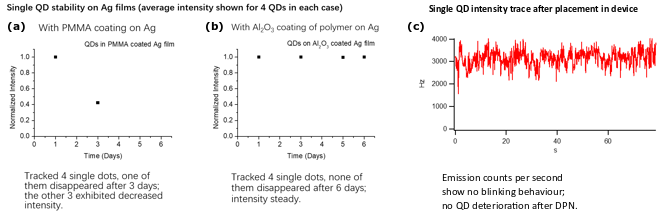}
    \caption{(a) Stability of QDs after being deposited onto polymer-coated Ag film from QD-suspensions in dichlorobenzene (DCB). (b) Stability of QDs after being deposited onto Al$_2$O$_3$-coated polymer-coated Ag film from QD-suspensions in DCB.  Taken together, results shown in (a) and (b) suggest that Al$_2$O$_3$ prevents the DCB solvent/QDs from penetrating the polymer layer, keeping the QDs separated from the metal. (c) Single gQDs remain non-blinking after deposition into device structures by DPN
    \label{Fig: stability}}
\end{figure}

\subsection*{Collection efficiency calculation}
To calculate the collection efficiency we first integrate over one ring (corresponding to a given value of the polar angle $\theta $) in the back focal plane image (Fig. \ref{Fig: estimation}c) : $\int_0^{2\pi} I(\theta,\phi) d\phi$ (Fig. \ref{Fig: estimation}d). 
As can be seen this distribution falls to zero at higher angles.
This enables us to extrapolate for the emission outside of our NA (from NA=0.9 to NA=1).
Now that we have estimated the overall intensity distribution we can conduct the integral over $\theta$ to obtain the collection efficiency (Fig. \ref{Fig: estimation}e) :
\begin{equation}
    \eta = \frac{\int_0^{\theta_{NA}}  d\theta \sin(\theta)  \int_0^{2\pi}d\phi I(\theta,\phi)}{\int_0^{\frac{\pi}{2}}d\theta \sin(\theta)\int_0^{2\pi}d\phi I(\theta,\phi)}
\end{equation}

\bibliography{mendeley}